# Deposition temperature dependence of thermo-spin and magneto-thermoelectric conversion in $Co_2MnGa$ films on $Y_3Fe_5O_{12}$ and $Gd_3Ga_5O_{12}$

Hayato Mizuno[1,a)], Rajkumar Modak[1], Takamasa Hirai[1], Atsushi Takahagi[2], Yuya Sakuraba[1], Ryo Iguchi[1], Ken-ichi Uchida[1,3,4,a)]

**AFFILIATIONS**

[1]National Institute for Materials Science, Tsukuba 305-0047, Japan

[2]Department of Mechanical Systems Engineering, Nagoya University, Nagoya 464-8601, Japan

[3]Institute for Materials Research, Tohoku University, Sendai 980-8577, Japan

[4]Center for Spintronics Research Network, Tohoku University, Sendai 980-8577, Japan

[a)]Author to whom correspondence should be addressed: MIZUNO.Hayato@nims.go.jp and UCHIDA.Kenichi@nims.go.jp

**ABSTRACT**

We have characterized $Co_2MnGa$ (CMG) Heusler alloy films grown on $Y_3Fe_5O_{12}$ (YIG) and $Gd_3Ga_5O_{12}$ (GGG) substrates at different deposition temperatures and investigated thermo-spin and magneto-thermoelectric conversion properties by means of a lock-in thermography technique. X-ray diffraction, magnetization, and electrical transport measurements show that the deposition at high substrate temperatures induces the crystallized structures of CMG while the resistivity of the CMG films on YIG (GGG) prepared at and above 500 °C (550 °C) becomes too high to measure the thermo-spin and magneto-thermoelectric effects due to large roughness, highlighting the difficulty of fabricating highly ordered continuous CMG films on garnet structures. Our lock-in thermography measurements show that the deposition at high substrate temperatures results in an increase in the current-induced temperature change for CMG/GGG and a decrease in that for CMG/YIG. The former indicates the enhancement of the anomalous Ettingshausen effect in CMG through crystallization. The latter can be explained by the superposition of the anomalous Ettingshausen effect and the spin Peltier effect induced by the positive (negative) charge-to-spin conversion for the amorphous (crystallized) CMG films. These results provide a hint to construct spin-caloritronic devices based on Heusler alloys.

Magnetic Heusler alloys have been widely studied as promising spintronic and spin-caloritronic materials because they may show half metallic[1,2] or topological band structure.[3] For instance, Co$_2$MnGa (CMG) with the $L2_1$ ordered phase is known to exhibit the large anomalous Hall and anomalous Nernst effects due to its topological nature.[4,5,6,7] Recent studies show that the CMG films with the $L2_1$ and $B2$ ordered phases also exhibit the large charge-spin current conversion efficiency.[8,9] This finding suggests that CMG would be useful for detecting and driving the thermo-spin effects such as the spin Seebeck[10,11] and spin Peltier effects (SPEs).[12,13,14] However, the spin-caloritronic effects in CMG have not been investigated except for the anomalous Nernst effect so far.

The SPE refers to the conversion of a spin current into a heat current in metal/magnetic-material junction systems. In combination with the charge-to-spin current conversion in the metal layer, e.g., the spin Hall effect (SHE),[15] the SPE is served as transverse thermoelectric conversion as follows. When the charge current is applied to the metal layer, it induces a transverse conduction electron spin current via the SHE and the conduction electron spin current is transformed into a magnon spin current in the attached magnetic layer via the spin-mixing conductance. The magnon spin current is eventually converted to a heat current and resultant temperature change via the SPE. Here, the directions of the input charge current and output heat current are perpendicular to each other. As shown in Figs. 1(a) and 1(b), in the in-plane magnetized configuration, the symmetry of the charge-to-heat current conversion due to the SHE-driven SPE is the same as that due to the anomalous Ettingshausen effect (AEE),[16] which is the Onsager reciprocal of the anomalous Nernst effect, in conductive ferromagnets. However, the SPE can be separated from the AEE by using a ferrimagnetic insulator, e.g., Y$_3$Fe$_5$O$_{12}$ (YIG), as the magnetic layer. Although the SPE has mainly been investigated in paramagnetic-metal (e.g., Pt)/YIG junction systems so far, the SPE-induced temperature change can appear also in ferromagnetic-metal/YIG junction systems because ferromagnetic metals exhibit the charge-to-spin conversion [see Fig. 1(c)].[17,18,19] The quantitative separation of the SPE from the AEE in the ferromagnetic-metal/YIG systems and the improvement of transverse thermoelectric conversion performance by the hybrid action of the SPE and AEE are important issues in spin caloritronics.

In this study, to clarify the potential of CMG as a spin-caloritronic material, we have grown CMG films on YIG and Gd$_3$Ga$_5$O$_{12}$ (GGG) substrates at different deposition temperatures and characterized the SPE and AEE by means of the lock-in thermography technique[13,20,21] as well as the structural, magnetic, and electrical transport properties. Here, GGG is employed as a reference substrate, which has the close lattice constant to and the same crystal structure as YIG. Since GGG is a paramagnetic insulator, it allows us to measure the AEE contribution in CMG, which is expected to be large because of the large anomalous Nernst effect of CMG,[4,6,7] free from the SPE contribution. This systematic dataset will provide a hint for future studies on spin-conversion and spin-caloritronic phenomena in ferromagnetic Heusler alloys, leading to the development of the spintronic thermal management.[22]

The sample systems used in this study are the 20-nm-thick CMG films deposited on single-crystalline YIG (111) and GGG (111) substrates by DC magnetron sputtering. We chose 20 nm because it is the reported minimum thickness of the ordered phases of CMG.[23] The YIG substrate consists of a 26-μm-thick YIG layer grown on a 0.5-mm-thick single-crystalline GGG (111) substrate by a liquid phase epitaxy method.[13] The lattice constant of CMG (YIG and GGG) is 5.8 (12.4) Å.[4,24] The YIG surface is mechanically polished with Al slurry. The dimension of the YIG and GGG substrates is $5 \times 10$ mm$^2$. The CMG films were sputtered from a $Co_{41.2}Mn_{27.5}Ga_{31.3}$ source with a base pressure of $< 2.0 \times 10^{-6}$ Pa, working pressure of 0.4 Pa in Ar atmosphere, sputtering power of 50 W, and the deposition rate of 0.06 nm/s. The substrate temperature ($T_{\text{sub}}$) during the deposition varied from room temperature to 550 °C. Before deposition, the YIG and GGG substrates were heated at 600 °C for 20 min to achieve a clean surface as previous studies performed for MgO substrates[25,26] and then down to $T_{\text{sub}}$ in the sputtering chamber. To avoid oxidation, the CMG layers were covered by the 2-nm-thick Al capping layers without breaking the vacuum, deposited at room temperature. The composition of the CMG films was determined to be $Co_{45.8}Mn_{26.2}Ga_{28.0}$ by x-ray fluorescence analysis (note that the composition of the sputtering source is intentionally tuned to obtain the optimized composition of the deposited films[7]). As a reference sample, we deposited the 20-nm-thick Pt film on the same YIG substrate at room temperature. The structural properties of the CMG films were characterized by out-of-plane x-ray diffraction. The magnetic properties were measured using a vibrating sample magnetometer. For electrical transport measurements, the films were patterned into Hall bars using photolithography and Ar-ion milling processes. The surface morphology of the films was evaluated by atomic force microscopy. The SPE and AEE were measured by means of the lock-in thermography method by patterning the CMG and Pt films into a U shape, the line width of which is 0.2 mm with the same process as the Hall bars. The procedures of the lock-in thermography measurements are the same as those shown in Ref. 13. All measurements were performed at room temperature.

Figures 2(a) and 2(b) show the x-ray diffraction patterns of the CMG/YIG and CMG/GGG systems, respectively. No clear peak of CMG appeared in the films deposited at room temperature, suggesting their amorphous structure. In contrast, the weak peaks of the CMG (220) lattice plane were observed for the high $T_{\text{sub}}$ deposited films. We confirmed that the 220 fundamental diffraction peak showed a ring shape in 2D scan, as shown in the inset of Fig. 2(b), indicating the polycrystalline structure with most fundamental *A*2 structure, where the *A*2 phase indicates the completely disordered structure among Co, Mn, and Ga. Although the presences of the *B*2 (partially disordered structure between Mn and Ga while Co atoms occupy regular sites) and $L2_1$ (fully ordered structure) phases can be recognized from 002 and 111 superlattice peaks, respectively, these peaks were too weak to be detected within the margin of experimental errors (note that the 002 peak was too weak even in the in-plane x-ray diffraction measurements). Therefore, the x-ray diffraction results observed here suggest the presence of polycrystalline structures but do not identify the ordered phases in the high $T_{\text{sub}}$ deposited CMG films on YIG and GGG.

Figure 2(c) shows the in-plane magnetic field *H* dependence of the magnetization *M* of the CMG films on the GGG substrates. We found that *M* saturates for $\mu_0 H > 0.04$ T, where $\mu_0$ is the vacuum permeability. The saturation magnetization $M_s$ of the high $T_{\text{sub}}$ deposited films was observed to be larger than that of the as-deposited ($T_{\text{sub}}$ = room temperature) film. The as-deposited film shows $\mu_0 M_s \sim 0.14$ T, which is much smaller than $\mu_0 M_s$ of the ordered CMG but in agreement with the previous reports.[27,28] The $\mu_0 M_s$ values for our high $T_{\text{sub}}$ deposited CMG films were estimated to be $\sim 0.45$ T, which is smaller than the reported values of the *A*2 phase ($\sim 0.54$ T)[27] and *B*2 ordered phase ($\sim 0.63$ T)[9,28] of CMG films. Figure 2(d) summarizes the measured and reference $M_s$ values.[8,9,27,28,29] This result suggests that our high $T_{\text{sub}}$ deposited CMG films consist of the mixture of the crystallized and amorphous structures.

We focus on the effect of $T_{\text{sub}}$ on electrical transport properties of the CMG films. Figure 2(e) shows the longitudinal electrical resistivity $\rho_{xx}$ as a function of $T_{\text{sub}}$. The $\rho_{xx}$ values of the CMG/YIG systems are larger than those of the CMG/GGG systems and their $T_{\text{sub}}$ dependence is different from each other. The resistivity of the CMG films on the YIG (GGG) substrates prepared at and above 500 °C (550 °C) was too high to apply a charge current (note that we prepared four CMG/YIG samples at $T_{\text{sub}}$ = 450 °C and two of them exhibit no electrical conduction while the remaining samples exhibit almost the same value of the electrical conductivity). Figure 2(f) shows the out-of-plane *H* dependence of the transverse resistivity $\rho_{xy}$ for the CMG/GGG systems. The $\rho_{xy}$ values show the *H*-odd dependence and its magnitude almost saturates at $\sim 0.7$ T for the high $T_{\text{sub}}$ deposited films. By extrapolating the $\rho_{xy}$ data above 1.0 T to zero field, the anomalous Hall resistivity for our high $T_{\text{sub}}$ deposited films was estimated to be $\sim 13$ μΩcm, which is comparable to the previously reported values for *B*2- and $L2_1$-ordered CMG films.[28,29] The large anomalous Hall resistivity implies that the high $T_\text{sub}$ deposited CMG/GGG films partially include the ordered phases. In order to clarify the reason for no electrical conduction for the high $T_\text{sub}$ deposited films, we investigated the surface morphology of the samples by atomic force microscopy. We confirmed that the average roughness $R_\text{a}$ of the CMG/YIG and CMG/GGG systems monotonically increases with increasing $T_\text{sub}$ and $R_\text{a}$ of CMG/YIG is larger than that of CMG/GGG at each $T_\text{sub}$ [Figs. 2(g) and 2(h)]. The large $R_\text{a}$ values for the samples prepared at high $T_\text{sub}$ result in the discontinuity of the CMG layers. Although the lattice constant and coefficient of thermal expansion of YIG and GGG are comparable,[24,30,31] the growth of CMG is sensitive to the surface energy, which determines the wettability of the initial growth of the CMG film; it is more difficult to obtain ordered CMG films with good electrical conduction when the YIG substrates are used than when the GGG substrates are used.

Next, we show the thermo-spin and thermoelectric conversion properties for the CMG/YIG and CMG/GGG systems with finite electrical conduction. During the lock-in thermography measurements, the square-wave-modulated AC charge current $J_\text{c}$ with the frequency $f = 5$ Hz and zero DC offset and in-plane $H$ were applied to the U-shaped CMG films, which is a standard condition for measuring the SPE and AEE.[13,16] By extracting the first-harmonic response of the current-induced temperature change, we can obtain lock-in amplitude $A$ and phase $\phi$ images respectively showing the amplitude and sign of the thermo-spin and/or thermoelectric conversion. To extract the pure SPE and AEE signals, we performed the lock-in thermography measurements with applying positive and negative $H$ and calculated the temperature modulation with the $H$-odd dependence:[32,33] $A_\text{odd}e^{-i\phi_\text{odd}} = \left(A(+H)e^{-i\phi(+H)} - A(-H)e^{-i\phi(-H)}\right)/2$, where the background due to the $H$-independent Peltier effect is excluded. The $A_\text{odd}$ and $\phi_\text{odd}$ images for the as-deposited CMG/YIG film are shown in Figs. 3(a) and 3(b), respectively, where $J_\text{c} = 3.0$ mA and $\mu_0|H| = 0.32$ T. We found that the clear current-induced temperature change appeared on the areas L and R of the CMG layer, where $\mathbf{J}_\text{c} \perp \mathbf{H}$, while no signal appeared in the area with $\mathbf{J}_\text{c} \parallel \mathbf{H}$. As shown in Fig. 3(b), $\phi_\text{odd}$ in L (R) is 0° (180°), indicating that the sign of the temperature modulation is reversed by reversing the $\mathbf{J}_\text{c}$ direction. Figure 3(c) [Fig. 3(d)] shows the increases of $A_\text{odd}$ in proportion to $J_\text{c}$ for all the CMG/YIG (CMG/GGG) systems for various values of $T_\text{sub}$. Figures 3(e) and 3(f) show the $\mu_0|H|$ dependence of $A_\text{odd}$ in the CMG/YIG and CMG/GGG systems, respectively. The magnitude of $A_\text{odd}$ is almost constant for $\mu_0|H| > 0.1$ T, where $M$ of the YIG and CMG layers is aligned along the $H$ direction. These results are consistent with the features of the temperature change induced by the SPE and AEE. We confirmed that all the samples exhibit the current-induced temperature modulation with the same sign.

In Fig. 4, we summarize the amplitude of the temperature modulation per unit charge current density, $A_{\mathrm{odd}}/j_{\mathrm{c}}$, where $j_{\mathrm{c}} = J_{\mathrm{c}}/(tw)$ with $t$ ($w$) being the thickness (line width) of the U-shaped CMG films. We also estimated $A_{\mathrm{odd}}/j_{\mathrm{c}}$ for the conventional Pt/YIG system to be $3.7 \times 10^{-13}$ Km$^2$A$^{-1}$, which is in good agreement with the previous reports[13,34] [see the star data point in Fig. 4(a)]. We found that, for the CMG/YIG systems, $A_{\mathrm{odd}}/j_{\mathrm{c}}$ in the as-deposited sample is 2.6 times larger than that in Pt/YIG and its magnitude is decreased in the high $T_{\mathrm{sub}}$ deposited samples [Fig. 4(a)]. In contrast, for the CMG/GGG systems, $A_{\mathrm{odd}}/j_{\mathrm{c}}$ is enhanced in the high $T_{\mathrm{sub}}$ deposited samples compared with the as-deposited sample [Fig. 4(b)]. For the as-deposited (high $T_{\mathrm{sub}}$ deposited) samples, $A_{\mathrm{odd}}/j_{\mathrm{c}}$ in the CMG/YIG systems is larger (smaller) than that in the CMG/GGG systems. The striking contrast in the current-induced temperature change between the CMG/YIG and CMG/GGG systems cannot be explained only by the AEE in the CMG layers, suggesting the coexistence of the SPE in the CMG/YIG systems.

Finally, we discuss the origin of the difference of $A_{\mathrm{odd}}/j_{\mathrm{c}}$ between the room temperature and high $T_{\mathrm{sub}}$ deposited films. In the CMG/GGG systems, the SPE does not contribute to the temperature modulation signals because GGG does not transport a spin current at room temperature.[35] Therefore, the signals for the CMG/GGG systems are due purely to the AEE in CMG. Since the deposition at high $T_{\mathrm{sub}}$ induces the crystallized structures of CMG, the increase in $A_{\mathrm{odd}}/j_{\mathrm{c}}$ between the CMG/GGG systems deposited at and above room temperature [Fig. 4(b)] indicates the enhancement of the AEE in CMG through crystallization. We also note that the magnitude of $A_{\mathrm{odd}}/j_{\mathrm{c}}$ is almost constant within the margin of experimental errors in the high $T_{\mathrm{sub}}$ deposited CMG/GGG systems, suggesting that the degree of ordering could be unchanged in the $T_{\mathrm{sub}}$ range, as suggested by almost constant $M_{\mathrm{s}}$ and $\rho_{xy}$ [Figs. 2(c) and 2(f)]. A similar trend has been reported in Ref. 36, where higher ordered phase exhibits a larger anomalous Nernst effect in a ferromagnetic Heusler alloy. Although the behaviors of the AEE in the CMG/YIG systems are expected to be similar to those in the CMG/GGG systems, the $A_{\mathrm{odd}}/j_{\mathrm{c}}$ signal in CMG/YIG is much larger than that in CMG/GGG in the as-deposited case and is decreased in the high $T_{\mathrm{sub}}$ deposited films [Fig. 4(a)]. This behavior can be explained by assuming that the SPE due to the charge-to-spin current conversion in the CMG layers in the as-deposited (high $T_{\mathrm{sub}}$ deposited) CMG/YIG systems makes an additive (subtractive) contribution to the AEE. This assumption indicates that the spin Hall angle of the amorphous (crystallized) CMG films is positive (negative) because the sign of the temperature modulation observed here is the same as that for the Pt/YIG systems with the positive spin Hall angle of Pt. Our scenario is consistent with the previous studies, which clarified the negative spin Hall angle of the *B*2-*L*2$_1$-mixed and *B*2-ordered CMG films.[8,9]

Although the SHE-driven SPE and AEE contributions in our crystallized CMG/YIG systems cancel each other out, the replacement of YIG by other ferrimagnets with the magnetization compensation can tune the sign of the SHE-driven SPE, enabling the enhanced thermoelectric conversion based on the hybrid action of the SPE and AEE.[37,38] The quantitative estimation of the spin Hall angle of the amorphous and crystallized CMG films on YIG remains to be performed. Nevertheless, our experiments reveal the fundamental behaviors of the SHE-driven SPE in the CMG/YIG systems and provide a possibility of tuning the sign and magnitude of the spin Hall angle of Heusler alloys through their crystallization.

In conclusion, we investigated the SPE and AEE in the CMG/YIG and CMG/GGG systems with the CMG films sputtered at different deposition temperatures. The CMG films grown on YIG at 500 and 550 °C and on GGG at 550 °C exhibit no electrical conduction due to the large roughness. These results indicate the difficulty of fabricating highly ordered continuous CMG films on garnet substrates. Further optimization of substrate and deposition conditions remains future work. Our lock-in thermography measurements show that the temperature change induced by the AEE in CMG is enhanced by the deposition at high $T_{\mathrm{sub}}$ through crystallization and that the SPE induces the additive (subtractive) temperature change due to the positive (negative) spin Hall angle of the amorphous (crystallized) CMG films. The systematic dataset reported here provides a guideline for future studies of spin caloritronics using magnetic Heusler alloys.

The authors thank H. Nagano for valuable discussions and M. Isomura, N. Kojima, and R. Tateishi for technical supports. This work was supported by the CREST "Creation of Innovative Core Technologies for Nano-enabled Thermal Management" (No. JPMJCR17I1) from JST, Japan; Grant-in-Aid for Scientific Research (S) (No. 18H05246) from JSPS KAKENHI, Japan; and the NEC Corporation. R.M. is supported by JSPS through the "JSPS Postdoctoral Fellowship for Research in Japan (Standard)" (P21064).

## AUTHOR DECLARATIONS
**Conflict of Interest**

The authors have no conflicts to disclose.

## DATA AVAILABILITY

The data that support the findings of this study are available from the corresponding author upon reasonable request.

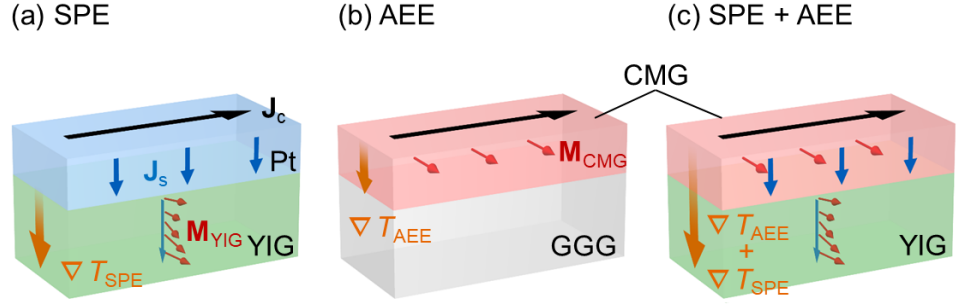

FIG. 1. Schematics of (a) the SPE in the Pt/YIG system, (b) the AEE in the CMG/GGG system, and (c) the hybrid thermoelectric conversion based on the SPE and AEE in the CMG/YIG system. $\mathbf{M}_{\text{YIG(CMG)}}$, $\mathbf{J}_{c(s)}$, and $\nabla T_{\text{SPE(AEE)}}$ represent the magnetization of YIG (CMG), charge current (spatial direction of the spin current), and temperature gradient induced by the SPE (AEE), respectively.

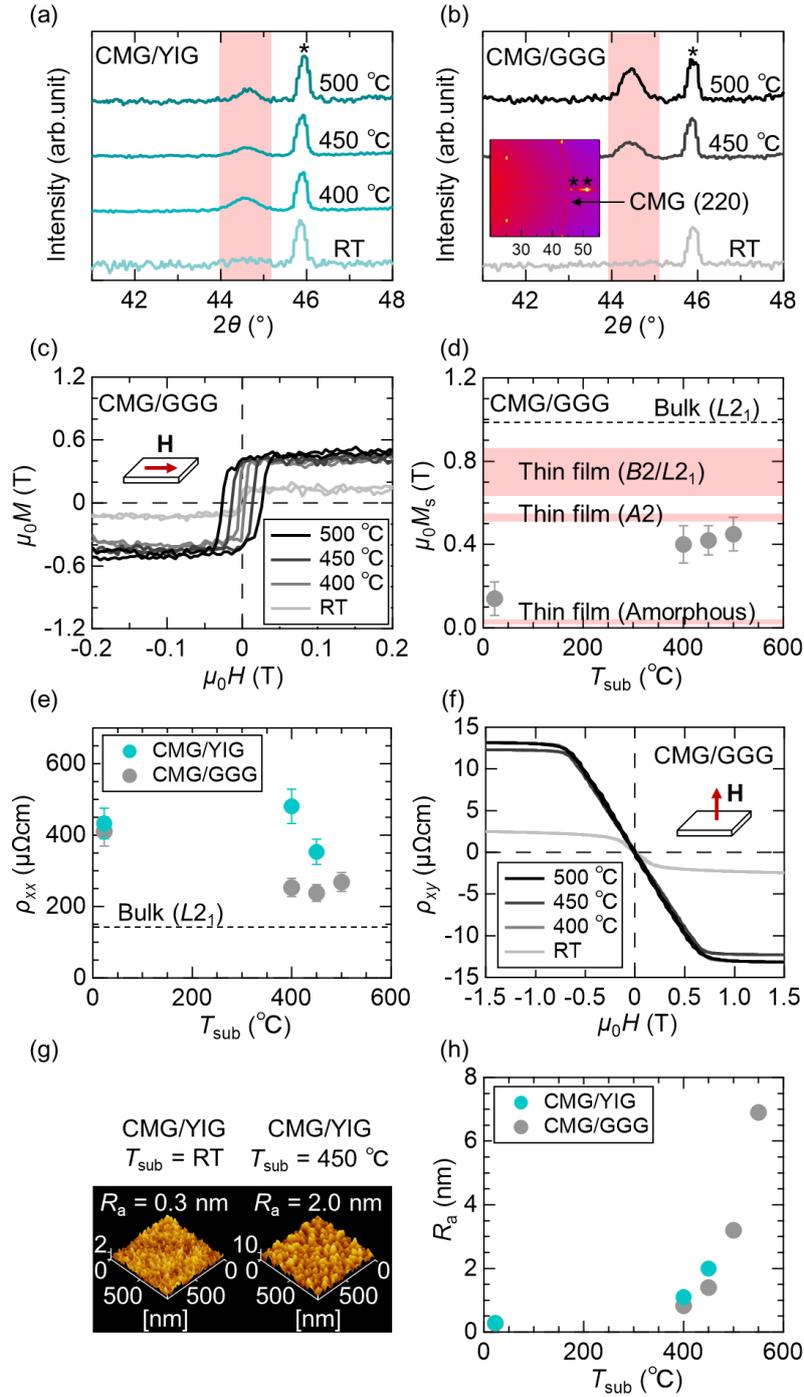

FIG. 2. (a) and (b) X-ray diffraction patterns of the CMG/YIG and CMG/GGG systems prepared at various values of the substrate temperature $T_{\mathrm{sub}}$. RT represents room temperature. The peaks highlighted in red arise from CMG (220). The peaks indicated by * arise from the YIG and GGG substrates. The inset shows the 2D scan for the CMG/GGG system prepared at $T_{\mathrm{sub}}$ = 500 °C. (c) In-plane $H$ dependence of the magnetization $M$ of the CMG/GGG systems prepared at various values of $T_{\mathrm{sub}}$. The paramagnetism contribution from the GGG substrates was subtracted by linear fitting of the raw data in the field range from $\mu_0|H|$ = 0.1 to 0.2 T, where $\mu_0$ is the vacuum permeability. (d) $T_{\mathrm{sub}}$

dependence of the saturation magnetization $M_s$ of the CMG films on the GGG substrates. The areas highlighted in red show the reference values of the amorphous[27,28], $A2$[27], $B2$[9,28,29], and $B2$-$L2_1$-mixed[8,28,29] structures of CMG films. (e) $T_\text{sub}$ dependence of the longitudinal electrical resistivity $\rho_{xx}$. (f) Out-of-plane $H$ dependence of the transverse resistivity $\rho_{xy}$ of the CMG films on the GGG substrates. (g) Atomic force microscope images of the CMG/YIG systems prepared at room temperature and 450 °C. (h) $T_\text{sub}$ dependence of the surface roughness $R_\text{a}$ of the CMG/YIG and CMG/GGG systems.

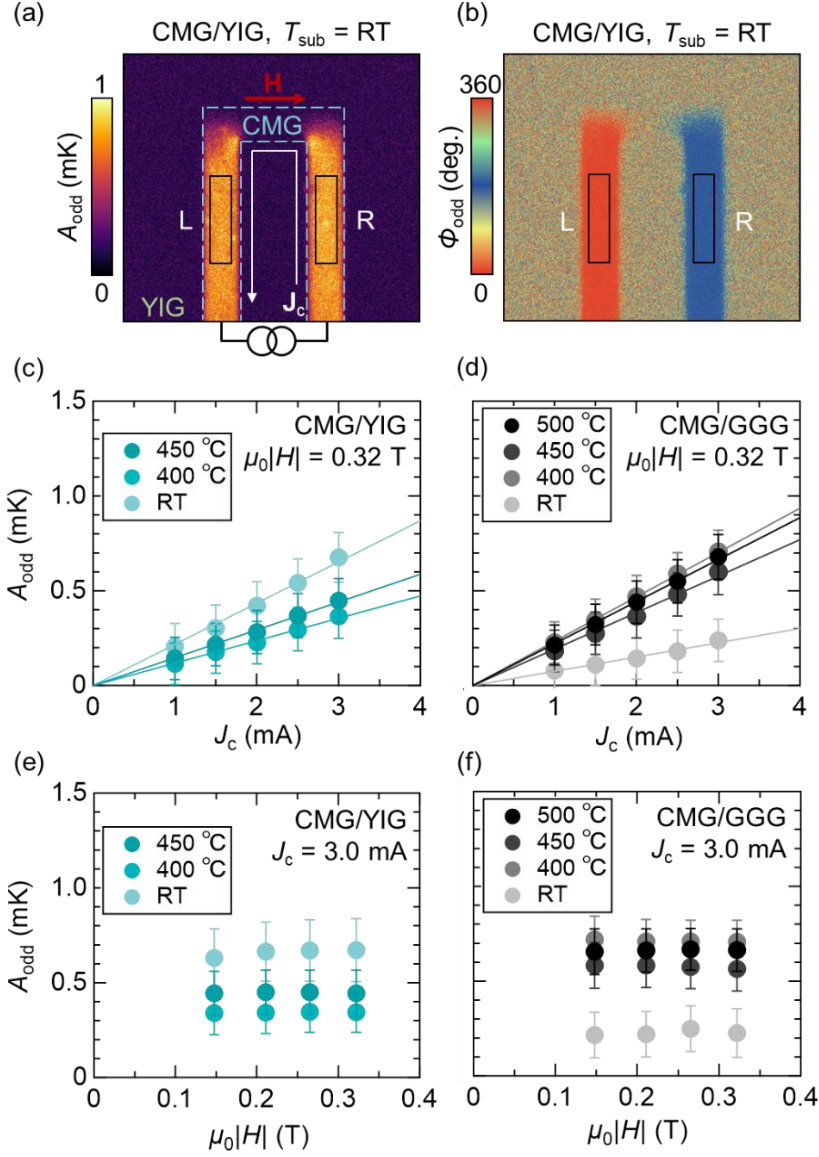

FIG. 3. (a) and (b) $A_{odd}$ and $\phi_{odd}$ images for the as-deposited CMG/YIG system at $J_c$ = 3.0 mA and in-plane field $\mu_0|H|$ = 0.32 T. $A_{odd}$ ($\phi_{odd}$) denotes the lock-in amplitude (phase) with the $H$-odd dependence. $J_c$ denotes the square-wave amplitude of the AC charge current applied to the U-shaped CMG layer. (c) and (d) $J_c$ dependence of $A_{odd}$ for the CMG/YIG and CMG/GGG systems at $\mu_0|H|$ = 0.32 T. The solid lines are obtained from the linear fitting. (e) and (f) $H$ dependence of $A_{odd}$ for the CMG/YIG and CMG/GGG systems at $J_c$ = 3.0 mA. The $A_{odd}$ data in (c)-(f) are extracted from the area R in (a).

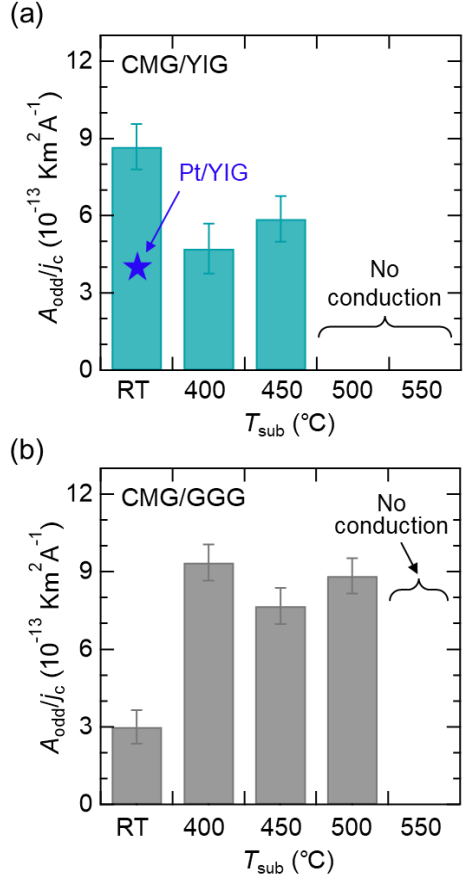

FIG. 4. $A_{\text{odd}}/j_c$ for the (a) CMG/YIG and (b) CMG/GGG systems for various values of $T_{\text{sub}}$. $j_c$ denotes the charge current density in the CMG layer. The $A_{\text{odd}}/j_c$ data are extracted from the slope of the linear fitting of the data in Figs. 3(c) and 3(d). The star data point indicates the $A_{\text{odd}}/j_c$ signal for the Pt/YIG system.